\documentclass[twocolumn,pra,aps]{revtex4}
\usepackage{graphicx, color, fancybox}
\usepackage{amsfonts}
\usepackage{amssymb,amsmath,amsthm}

\begin{document}
\title{Complete solution for unambiguous discrimination of three pure states with 
real inner products}
\author{H.~Sugimoto, T.~Hashimoto, M.~Horibe, and A.~Hayashi}
\affiliation{Department of Applied Physics, 
           University of Fukui, Fukui 910-8507, Japan}

\begin{abstract}
Complete solutions are given in a closed analytic form for unambiguous discrimination 
of three general pure states with real mutual inner products. 
For this purpose, we first establish some general results on unambiguous discrimination of $n$ linearly independent pure states. The uniqueness of solution is 
proved. The condition under which the problem is reduced to an $(n-1)$-state problem 
is clarified. After giving the solution for three pure states with real mutual 
inner products, we examine some difficulties in extending our method to the case of complex 
inner products. There is a class of set of three pure states with complex 
inner products for which we obtain an analytical solution. 
\end{abstract}

\maketitle

\newcommand{\ket}[1]{|\,{#1}\,\rangle}
\newcommand{\bra}[1]{\langle\,{#1}\,|}
\newcommand{\braket}[2]{\langle\,#1\,|\,#2\,\rangle}
\newcommand{\mbold}[1]{\mbox{\boldmath $#1$}}
\newcommand{\sbold}[1]{\mbox{\boldmath ${\scriptstyle #1}$}}
\newcommand{\tr}{\,{\rm tr}\,}
\newcommand{\trm}{{\rm tr}}

\newcommand{\rmR}{{\rm R}}
\newcommand{\doubleket}[1]{|\,{#1}\,\rangle\rangle}
\newcommand{\doublebra}[1]{\langle\langle\,{#1}\,|}
\newcommand{\doublebraket}[2]{\langle\langle\,#1\,|\,#2\,\rangle\rangle}

\newtheorem{theorem}{Proposition}
\renewcommand{\qedsymbol}{\rule[0pt]{1.5ex}{1.5ex}}

\section{Introduction} 
In quantum mechanics, there is no way to distinguish nonorthogonal states perfectly. 
This is because quantum measurement is statistical in nature, and it generally 
destroys the state of the system to be measured. Considerable works have been done 
to find optimal ways to distinguish quantum states in various situations 
\cite{Helstrom76, Holevo82, Chefles00}.  
These studies are stimulated by needs of quantum communication and 
quantum cryptography where the discrimination of quantum
states is one of the key issues. 

In the quantum state discrimination problem, two settings have been commonly 
investigated. In minimum-error discrimination \cite{Helstrom76}, 
the discrimination success probability is maximized without any constraint on the 
probability of erroneous results. In unambiguous discrimination, however, the success 
probability is maximized under the condition that measurement should not produce 
erroneous results. This is possible by allowing an inconclusive result: ``I don't know.'' 
Ivanovic \cite{Ivanovic87}, Dieks \cite{Dieks88}, and Peres \cite{Peres88} 
found the optimal solution for unambiguous discrimination of two pure states with 
equal occurrence probabilities. Later the case of general occurrence probabilities 
was solved in Ref.~\cite{Jaeger95}. Unambiguous discrimination was experimentally 
demonstrated in Ref.~\cite{Clarke01}. 
A scheme that interpolates the two discrimination settings has also been 
proposed, and its solution for two general pure states was given in 
Refs.~\cite{Hayashi08,Sugimoto09}. 

For unambiguous discrimination among more than two pure states, it is not easy to 
obtain analytical solutions, though a great deal of works on general theories have 
been reported: For example see Refs.~\cite{Chefles98,Chefles_Barnett98,Peres98, 
Sun01,Zhang01,Eldar_IEEE03,Jafarizadeh08,Samsonov09,Pang09}. 
The same thing is true for unambiguous discrimination for mixed states 
\cite{Rudolph03,Raynal03,Eldar04,Feng04,Herzog04,Herzog07,Zhou07,Kleinmann10}. 

The case of three pure states has also been considered as an application by some articles dealing with general theories. Even for three pure states, to the best of our knowledge, complete analytical results 
are known only in limited special cases: 
(i) The case of symmetric states with equal occurrence probabilities \cite{Chefles_Barnett98}, in which the states are generated 
by a single unitary operator; 
(ii) the case in which two of the three real mutual inner products are equal and equal occurrence probabilities are assumed \cite{Sun01}; and (iii) the case of three pure states with general occurrence probabilities and one of the mutual 
inner products being 0 \cite{Pang09}.   

The purpose of this article is to give complete analytical solutions for 
unambiguous discrimination of three pure states with real mutual inner products 
and general occurrence probabilities. In unambiguous discrimination, 
the optimal strategy may produce a vanishing probability of identifying the input 
state with some of the states. This issue will be thoroughly analyzed. 

In Sec.~\ref{sec_General}, we first establish some general results on unambiguous discrimination of $n$ pure states. We formulate the problem as semidefinite programming. 
It turns out very useful to consider a problem in which some constraints in the problem
are relaxed. This relaxed problem is used to clarify the conditions under which some identification probabilities vanish, and the problem is reduced to a certain 
$(n-1)$-state problem. 
Then, complete analytical solutions of three pure states with real mutual inner products 
are presented in Sec.~\ref{sec_Three}. 
In Sec.~\ref{sec_Discussion}, we discuss some difficulties 
in extending our method to the case of general complex inner products. 
In the appendix, we show that solutions of unambiguous discrimination 
for linearly independent $n$ pure states are generally unique.

\section{General consideration} \label{sec_General}
\subsection{Formulation of problem}
Suppose that a state is drawn from a set of known $n$ pure states $\ket{\phi_i}$ with 
probabilities $\eta_i$. We assume that the $n$ states are linearly independent and 
the occurrence probabilities $\eta_i$ are nonzero. The task is 
to unambiguously identify the given state with one in the set of $n$ states: 
An erroneous identification is not allowed, but an inconclusive result 
(``I don't know'') is possible. 

Measurement is described by a positive operator-valued measure (POVM), and it consists 
of $n+1$ elements $\{E_1,E_2,\ldots,E_n,E_?\}$, each element being a positive 
semidefinite operator on the $n$-dimensional space $V$ spanned by $\ket{\phi_i}$.  
When the outcome is $i\ (i=1,\ldots,n)$, the given input state is identified with 
the state $\ket{\phi_i}$. The element $E_?$ corresponds to the inconclusive result. 
The POVM satisfies the following completeness relation:
\begin{align}
  E_? + \sum_{i=1}^n E_i = \mbold{1}_V. \label{eq:completeness}
\end{align} 

The no-error conditions are written as 
\begin{align}
  \bra{\phi_j} E_i \ket{\phi_j} = 0\ \ (i \ne j), \label{eq:no-error-condition}
\end{align}
which require that the support of the positive semidefinite operator $E_i$ be  
the one-dimensional subspace that is orthogonally complement to the $(n-1)$-dimensional 
subspace spanned by the states $\{ \ket{\phi_j} \}_{j(\ne i)}$. 
Therefore, $E_i$ should take the following form \cite{Chefles98}:  
\begin{align}
  E_i = x_i \ket{\tilde \phi_i}\bra{\tilde \phi_i},  
\end{align}
where $x_i$ is a non-negative constant to be determined, and the state 
$\ket{\tilde \phi_i}$ is orthogonal to any state $\ket{\phi_j}$ for $j \ne i$. 
For convenience we normalize the states $\ket{\tilde \phi_i}$ such that 
\begin{align}
   \braket{\tilde \phi_i}{\phi_j} = \delta_{ij}.
\end{align} 
Since the states $\ket{\phi_i}$ are linearly independent, those states 
$\ket{\tilde \phi_i}$ are uniquely given by 
\begin{align}
   \ket{\tilde \phi_i} = \sum_{j=1}^n (N^{-1})_{ji} \ket{\phi_j}, 
\end{align}
where
\begin{align*}
N_{ij}\equiv\braket{\phi_i}{\phi_j}, 
\end{align*}
is the Gram matrix of the 
states to be discriminated. 

The coefficient $x_i$ represents the conditional probability 
of successfully identifying an input state given that the input state is 
$\ket{\phi_i}$. In terms of $x_i$, the discrimination 
success probability is given by 
\begin{align}
  p = \sum_{i=1}^n \eta_i \bra{\phi_i}E_i\ket{\phi_i} = \sum_{i=1}^n \eta_i x_i. 
\end{align}

Now let us examine what condition the positivity of $E_?$ further imposes on the  coefficients $x_i$. By the completeness relation (\ref{eq:completeness}), 
the positivity of $E_?$ is expressed as 
\begin{align}
  \mbold{1}_V - \sum_{i=1}^n x_i \ket{\tilde \phi_i}\bra{\tilde \phi_i} \ge 0. 
                     \label{eq:e0-positivity}
\end{align}
This condition is equivalent to the positivity of the $n \times n$ matrix $M$ defined by 
\begin{align}
  M_{ij} &= \bra{\phi_i} 
     \left( \mbold{1}_V - \sum_{k=1}^n x_k \ket{\tilde \phi_k}\bra{\tilde \phi_k} \right)
           \ket{\phi_j} \nonumber \\
         &= N_{ij} - x_i \delta_{ij},  
\end{align}
since $\ket{\phi_i}$ are linearly independent.  

Thus, the unambiguous discrimination problem of the ensemble 
$\{\eta_i, \ket{\phi_i}\}_{i=1}^{n}$ is formulated as semidefinite programming: 
The problem is to maximize 
\begin{subequations} 
   \label{eq:n-state-problem}
\begin{align}
  p = \sum_{i=1}^n \eta_i x_i, 
\end{align}
subject to conditions given by 
\begin{align}
  & x_i \ge 0\ \ (i=1,2,\ldots,n) , \label{eq:positive-x} \\
  & N-X \ge 0, \label{eq:N-X}
\end{align}
\end{subequations}
where $X=\text{diag}(x_1,x_2,\ldots,x_n)$ and 
variables are $x_i$. We can show that solutions $x_i$ are uniquely 
determined; the optimal measurement is unique. The proof is presented in the appendix. 

\subsection{Relaxed problem and reduction of the number of states} 
In unambiguous discrimination, the optimal strategy may produce
a vanishing probability of identifying the input state with some of the states 
$\ket{\phi_i}$: Some of $x_i$ may be zero in solutions of the problem  
(\ref{eq:n-state-problem}). 
To see when this happens, it is convenient to 
consider the problem where the conditions (\ref{eq:positive-x}), $x_i \ge 0$, 
are omitted. To distinguish the two problems, this relaxed form of the 
problem is called a ``relaxed $n$-state problem'' from now on, and the original one  
(\ref{eq:n-state-problem}) is simply called an ``$n$-state problem.'' 

In the relaxed $n$-state problem, the task is to maximize 
\begin{subequations} 
   \label{eq:relaxed-n-state-problem}
\begin{align}
  p^\rmR = \sum_{i=1}^n \eta_i x_i^\rmR, 
\end{align}
subject to conditions given by 
\begin{align}
  N-X^\rmR \ge 0, \label{eq:relaxed-condition}
\end{align}
\end{subequations}
where $X^\rmR=\text{diag}(x_1^\rmR,x_2^\rmR,\ldots,x_n^\rmR)$. 
We use a superscript $\rmR$ to represent quantities in the 
relaxed problem.  

There are important relations between solutions of the two problems.
Let $\mbold{x}^\rmR=(x_1^\rmR,x_2^\rmR,\ldots,x_n^\rmR)$ be a solution 
of the relaxed $n$-state problem.  Then we can show the following properties: 
\begin{description}
\item[(I)] If $x_i^\rmR \ge 0$ for all $i$, then $\mbold{x}^\rmR$ is the  
solution of the $n$-state problem. 
\item[(II)] If $x_i^\rmR < 0$ for $i$ in some nonempty index set $S$, then 
the solution of the $n$-state problem has vanishing components, $x_i =0$, for 
at least one $i$ in $S$.  
\end{description}
Property (I) is evident. To show property (II), we consider 
a linear interpolation between the two solutions as 
\begin{align}
 \mbold{\hat x}(\kappa) = (1-\kappa) \mbold{x} + \kappa \mbold{x}^\rmR,\ \ 0 \le \kappa \le 1, 
\end{align}
where $\mbold{x}$ is the solution of the $n$-state problem and $\mbold{x}^\rmR$ is a 
solution of the relaxed $n$-state problem. We find that $\mbold{\hat x}(\kappa)$ satisfies the 
condition (\ref{eq:N-X}) since  
\begin{align}
  N-\hat X(\kappa) = (1-\kappa)(N-X) + \kappa (N-X^\rmR) \ge 0. 
\end{align}
The success probability corresponding to $\mbold{\hat x}(\kappa)$ is written as 
\begin{align}
  \hat p(\kappa) = (1-\kappa)p_{\max} + \kappa p_{\max}^\rmR,
\end{align}
where $p_{\max}$ and $p_{\max}^\rmR$ are the maximum success probabilities of the 
$n$-state and relaxed $n$-state problems, respectively. 
Note that $p_{\max}^\rmR \ge p_{\max}$ by definition. 
From this we note that $\hat p(\kappa)$ is no less than $p_{\max}$ for 
$0 \le \kappa \le 1$. This implies that $\mbold{\hat x}(\kappa)$ is a solution 
of the $n$-state problem if its all components $\hat x_i(\kappa)$ are nonnegative. 
%
%

\begin{figure}
\includegraphics[width=7cm]{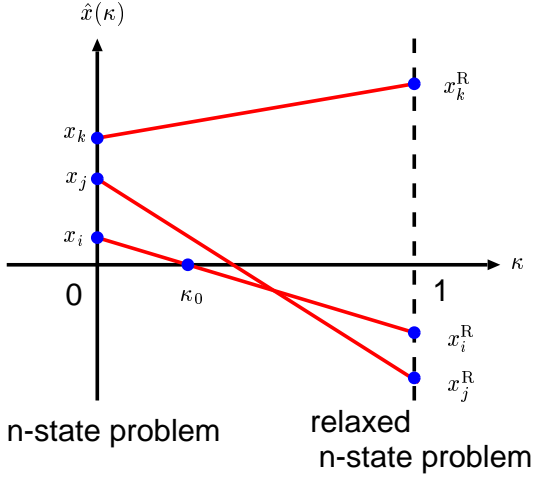}
\caption{\label{fig:solution}
(Color online) Linear interpolation of the solutions of the $n$-state problem and the 
relaxed $n$-state problem. Each component of $\mbold{\hat x}(\kappa)$ is plotted vs. 
$\kappa$. There exists $ 0 \le \kappa_0 < 1$ such that $\hat x_k(\kappa_0) \ge 0$ 
for all $k$  and $\hat x_i(\kappa_0) = 0$ for a certain $i \in S$. 
By the uniqueness of the solution of the $n$-state problem, 
$\kappa_0$ is actually 0. 
}
\end{figure}

Obviously, $\hat x_i(\kappa)$ is continuous with respect to $\kappa$.
We also know that $x_i \ge 0$ for all $i$ and that $x_i^\rmR < 0$ for $i \in S$, 
whereas $x_i^\rmR \ge 0$ otherwise.
Using the intermediate value theorem, we observe that there exists 
$ 0 \le \kappa_0 < 1$ such that $\hat x_k(\kappa_0) \ge 0$ for all $k$  and 
$\hat x_i(\kappa_0) = 0$ for a certain $i \in S$ (see Fig. \ref{fig:solution}).
Therefore, $\mbold{\hat x}(\kappa_0)$ is a solution for the $n$-state problem. 
Remember that the solution of the $n$-state problem is unique, implying $\kappa_0 = 0$. 
Thus, we can conclude that the solution of the $n$-state 
problem has vanishing components, $x_i =0$, for at least one $i$ in $S$. 
This completes the proof of property (II). 

Next, we show that if some of $x_i^\rmR$ are negative in a solution of the relaxed 
$n$-state problem, the $n$-state problem can be reduced to a certain $(n-1)$-state problem. 
Let us first assume that only one component is negative.  Without 
loss of generality, we assume that $x_n^\rmR < 0$ and the others are non-negative.  
Property (II) then states that the component $x_n$ is 0 in the solution of the $n$-state 
problem. This solution can be obtained by requiring an additional 
constraint, $x_n=0$, to the original $n$-state problem. With this additional 
constraint, the condition (\ref{eq:e0-positivity}) becomes 
\begin{align}
  \mbold{1}_V - \sum_{i=1}^{n-1} x_i \ket{\tilde \phi_i}\bra{\tilde \phi_i} \ge 0. 
                     \label{eq:e0-positivity-prime1}
\end{align} 
Let us define the following two sets of $n-1$ states:
\begin{align}
 &\ket{\phi_i'} \equiv \frac{Q \ket{\phi_i}}
                            {\sqrt{\bra{\phi_i} Q \ket{\phi_i}}}\ \ 
                      (i=1,2,\ldots,n-1),  \label{eq:phi-prime} \\ 
 &\ket{\tilde\phi_i'} \equiv \sqrt{\bra{\phi_i} Q \ket{\phi_i}} \ket{\tilde\phi_i}\ \
                      (i=1,2,\ldots,n-1),  \label{eq:tilde-phi-prime} 
\end{align}
where  
\begin{align}
    Q \equiv \mbold{1}_V - \ket{\phi_n}\bra{\phi_n}, 
\end{align}
is the projector onto the $(n-1)$-dimensional subspace $V'$ spanned by 
the $n-1$ states $\{ \ket{\tilde \phi_i} \}_{i=1}^{n-1}$. 

It is readily verified that the two sets of states are reciprocal in the sense that 
$\braket{\tilde \phi_i'}{\phi_j'} = \delta_{ij}$. Furthermore, the condition 
(\ref{eq:e0-positivity-prime1}) can be expressed as 
\begin{align}
  \mbold{1}_{V'} - \sum_{i=1}^{n-1} x_i' \ket{\tilde \phi_i'}\bra{\tilde \phi_i'} \ge 0, 
                     \label{eq:e0-positivity-prime2}
\end{align}
where $x_i'$ is defined as 
\begin{align}
   x_i' \equiv \frac{x_i}{\bra{\phi_i} Q \ket{\phi_i}}. 
\end{align}
We will shortly see that $x_i'$ are variables in the reduced $(n-1)$-state problem. 
The success probability of the $n$-state problem with the additional constraint, 
$x_n=0$, is rewritten as 
\begin{align} 
   p &= \sum_{i=1}^{n-1} \eta_i x_i  \nonumber \\
     &= \left (\sum_{k=1}^{n-1} \eta_k \bra{\phi_k} Q \ket{\phi_k} \right)
         \left ( \sum_{i=1}^{n-1} \eta_i' x_i' \right), 
             \label{eq:max_probability_reduced}
\end{align} 
where $\eta_i'$ is defined as
\begin{align}
    \eta_i' \equiv \frac{\eta_i \bra{\phi_i} Q \ket{\phi_i} }
                        { \sum_{k=1}^{n-1} \eta_k \bra{\phi_k} Q \ket{\phi_k} }.
                                   \label{eq:eta-prime}
\end{align}
Note that $\eta_i'>0$ and $\sum_{i=1}^{n-1}\eta_i'=1$. The $\eta_i'$ turn out 
to be the occurrence probabilities in the $(n-1)$-state problem. 
From Eq. (\ref{eq:max_probability_reduced}) and the condition 
(\ref{eq:e0-positivity-prime2}), we observe that the $n$-state problem 
with the additional constraint $x_n = 0$ is equivalent to the 
$(n-1)$-state problem for the ensemble $\{ \eta_i',\,\ket{\phi_i'} \}_{i=1}^{n-1}$. 
The maximum success probabilities and the variables $x_i$ of the 
two problems are related in the following way:
\begin{align}
 & p_{\max}^{(n)}\left[  \{\eta_i,\ket{\phi_i}\}_{i=1}^{n} \right]  \nonumber \\
 & = \left (\sum_{k=1}^{n-1} \eta_k \bra{\phi_k} Q \ket{\phi_k} \right)
                   p_{\max}^{(n-1)} 
                    \left[ \{\eta_i',\ket{\phi_i'}\}_{i=1}^{n-1} \right], 
                        \label{eq:p-max-reduction} \\
 & x_i = \bra{\phi_i} Q \ket{\phi_i} x_i'.     \label{eq:x-i-reduction}               
\end{align}
It should be emphasized that the extra condition $x_n=0$ does not imply that 
the state $\ket{\phi_n}$ can completely be neglected. This is because the no-error 
conditions (\ref{eq:no-error-condition}) should be respected for all the states including 
$\ket{\phi_n}$. The states to be discriminated in the reduced $(n-1)$-state problem 
are not $\{\ket{\phi_i}\}_{i=1}^{n-1}$ but $\{\ket{\phi_i'}\}_{i=1}^{n-1}$, which are 
the normalized states with their $\ket{\phi_n}$ component subtracted. 
As can be seen from (\ref{eq:eta-prime}), the occurrence probabilities are also modified 
by the factor of $\bra{\phi_i} Q \ket{\phi_i}$. 

So far, we assumed that only one component of $\mbold{x}^\rmR$ is negative. 
If more than one component is negative, 
we do not know which $x_i\ (i \in S)$ vanishes in the $n$-state problem. 
Therefore, for each $i$ in $S$, we need to solve the corresponding $(n-1)$-state problem 
with $x_i$ being set 0. From Eq.~(\ref{eq:p-max-reduction}), we obtain 
a finite number of candidates for the maximum success probability of the 
$n$-state problem. Evidently, the one with the largest probability is the solution 
of the $n$-state problem. 

We can show that the number of negative components of $\mbold{x}^\rmR$ is 
$n-1$ at most. This can be seen in the following way: 
The Gram matrix $N$ is strictly positive definite since the states 
$\{\ket{\phi_i}\}_{i=1}^n$ are linearly independent. This implies that 
the condition $N-X^{\rmR} \ge 0 $ is satisfied for a set of sufficiently small but 
positive $x_i^{\rmR}$, which leads to a positive success probability. Therefore, $\mbold{x}^\rmR$ with all components negative cannot be a solution of the relaxed 
$n$-state problem.  

\subsection{Dual problem of the relaxed problem} 
In our method, we first solve the relaxed $n$-state problem defined in 
Eq.~(\ref{eq:relaxed-n-state-problem}). 
If all $x_i^\rmR$ are non-negative, they give the solution of the $n$-state problem. 
As shown in the preceding section, if some of $x_i^\rmR$ are negative, the 
problem is reduced to a certain $(n-1)$-state problem. How, then, do we solve the relaxed 
$n$-state problem? We find it convenient to consider the dual problem of the  relaxed $n$-state problem. 

The task in the dual problem is to minimize 
\begin{subequations} 
   \label{eq:dual-problem}
\begin{align}
   d \equiv \tr YN ,
\end{align}
subject to the conditions given by 
\begin{align}
   Y \ge 0,\ Y_{ii}=\eta_i , \label{eq:dual-condition}
\end{align}
\end{subequations}
where the variable is an $n \times n$ Hermitian matrix $Y$. 
It is readily shown that the minimum $d_{\min}$ gives an upper bound for 
the maximum $p_{\max}^\rmR$ in the relaxed $n$-state problem (\ref{eq:relaxed-n-state-problem}). 
\begin{align}
  p_{\max}^\rmR &= \sum_{i=1}^n \eta_i x_i^\rmR  = \sum_{n-1}^n Y_{ii}x_i^\rmR  
                                             \nonumber \\
       &= \tr YX^\rmR = \tr Y [N-(N-X^\rmR) ]     \nonumber \\
       &\le \tr YN =d_{\min}, 
\end{align}       
where the inequality in the last line follows from the inequality $\tr Y(N-X^\rmR) \ge 0$. 
Therefore, the set of conditions given by 
\begin{align}
   (N-X^\rmR)Y = 0, \label{eq:attainability}
\end{align}
together with the conditions (\ref{eq:relaxed-condition}) and (\ref{eq:dual-condition}) 
is a sufficient condition for $d_{\min}$ to be equal to $p_{\max}^{\rmR}$.  
In the next section, for the $n=2$ and $n=3$ cases, we first solve the dual problem (\ref{eq:dual-problem}) 
and then we explicitly verify that this set of conditions 
are satisfied. According to the general theory of semidefinite programming 
\cite{Vandenberghe96}, it is guaranteed that the solution of the $n$-state problem can 
always be obtained in this way.  

The variable matrix $Y$ in the dual problem is constrained by the condition 
(\ref{eq:dual-condition}). Any positive semidefinite matrix can be expressed as the Gram matrix 
of some set of vectors. Therefore, we can parametrize $Y$ satisfying the condition 
(\ref{eq:dual-condition}) as 
\begin{align}
   Y_{ij} = \braket{y_i}{y_j}, \label{eq:Y-parameterization} 
\end{align} 
where $\ket{y_i}$ are arbitrary vectors with $\big| \ket{y_i} \big|=\sqrt{\eta_i}$. 
This convenient parametrization will be used 
in the following sections.     

\section{Three pure states with real inner products} 
\label{sec_Three}
In our scheme, we solve the dual problem (\ref{eq:dual-problem}) to obtain 
the solution of the relaxed problem (\ref{eq:relaxed-n-state-problem}). 
If all $x_i^\rmR$ are non-negative, they give the solution of the 
original nonrelaxed problem (\ref{eq:n-state-problem}). Otherwise, the problem is 
reduced to the one for $n-1$ states.  To demonstrate how this scheme works, 
we first solve the problem of two pure states by our scheme. 
This provides a helpful guideline for the three-state problem which will 
be discussed next. 

\subsection{Two pure states}\label{sec:n_2}
The task of the dual problem is to minimize
\begin{eqnarray}
d = \tr{YN}, \label{eq:dual_d_of_two_state}
\end{eqnarray}
subject to
\begin{eqnarray}
Y \geq 0, \ Y_{ii}=\eta_{i}, 
\end{eqnarray}
where variable $Y$ is a 2$\times$2 Hermitian matrix. 
By changing the phase of the state $\ket{\phi_{1}}$, we can assume that the mutual  
inner product is real and negative, $\braket{\phi_{1}}{\phi_{2}}=-\left|\braket{\phi_{1}}{\phi_{2}}\right|$. 
Using the parametrization (\ref{eq:Y-parameterization}), we can write $d$ as 
\begin{equation}
d= 1-\left(\braket{y_{1}}{y_{2}}+\braket{y_{2}}{y_{1}}\right)
   \left|\braket{\phi_{1}}{\phi_{2}}\right|. \label{eq:minimizing_d_of_two_state}
\end{equation}
Applying the triangle inequality and Schwarz's inequality to the second term of 
Eq.~(\ref{eq:minimizing_d_of_two_state}), we have 
\begin{align*}
\left|\braket{y_{1}}{y_{2}}+\braket{y_{2}}{y_{1}}\right| 
   &\leq 2\left|\braket{y_{1}}{y_{2}}\right|  \\
   &\leq 2\sqrt{\braket{y_{1}}{y_{1}}}\sqrt{\braket{y_{2}}{y_{2}}} \\
   &= 2\sqrt{\eta_{1}\eta_{2}},
\end{align*}
where $\braket{y_{i}}{y_{i}}=\eta_{i}$ is used. 
By this inequality, we obtain the minimum of $d$ to be 
\begin{eqnarray}
d_{\min} = 1-2\sqrt{\eta_{1}\eta_{2}}\left|\braket{\phi_{1}}{\phi_{2}}\right|, 
\end{eqnarray}
and the vectors $\ket{y_{i}}$ attaining the minimum can be taken to be 
the following one-dimensional vectors: 
\begin{align*}
\ket{y_{1}}=\sqrt{\eta_{1}},\hspace{1ex}\ket{y_{2}}=\sqrt{\eta_{2}}.
\end{align*}
Thus, $Y$ is given by
\begin{eqnarray}
Y = \left(
\begin{array}{cc}
\eta_{1}&\sqrt{\eta_{1}\eta_{2}}\\
\sqrt{\eta_{2}\eta_{1}}&\eta_{2}
\end{array}
\right) 
= \mbold{\beta}\mbold{\beta}^{\dagger}, \label{eq:Y_of_two_state}
\end{eqnarray} 
with $\mbold{\beta}$ defined to be 
\begin{eqnarray*}
\mbold{\beta}\equiv\left(
\begin{array}{c}
\sqrt{\eta_{1}}\\
\sqrt{\eta_{2}}
\end{array}
\right). 
\end{eqnarray*} 
 
The attainability condition is given by Eq.~(\ref{eq:attainability}). Substituting 
$Y$ in Eq.(\ref{eq:attainability}) and multiplying this 
equation by $\mbold{\beta}$ on the right, we obtain
\begin{eqnarray}
\left(N-X^\rmR \right)\mbold{\beta}=0, \label{eq:attainability_condition_of_two_state_by_beta}
\end{eqnarray}
since $\mbold{\beta}^{\dagger}\mbold{\beta}=1$. 
By using the matrix representation, 
Eq.~(\ref{eq:attainability_condition_of_two_state_by_beta}) is rewritten as
\begin{eqnarray*}
\left(
\begin{array}{cc}
1-x^{\rm R}_{1}&-\left|\braket{\phi_{1}}{\phi_{2}}\right|\\
-\left|\braket{\phi_{1}}{\phi_{2}}\right|&1-x^{\rm R}_{2}
\end{array}
\right)\left(
\begin{array}{c}
\sqrt{\eta_{1}}\\
\sqrt{\eta_{2}}
\end{array}
\right) = 0. 
\end{eqnarray*}
From this equation, we immediately find that $x^{\rm R}_{1}$ and $x^{\rm R}_{2}$ are 
given by 
\begin{align*}
x^{\rm R}_{1}&=1-\sqrt{\frac{\eta_{2}}{\eta_{1}}}\left|\braket{\phi_{1}}{\phi_{2}}\right|, \\
x^{\rm R}_{2}&=1-\sqrt{\frac{\eta_{1}}{\eta_{2}}}\left|\braket{\phi_{1}}{\phi_{2}}\right|. 
\end{align*}

The $x^{\rm R}_{i}$ must further satisfy Eq.(\ref{eq:relaxed-condition}): The eigenvalues 
of $N-X^{\rm R}$ should be non-negative. The characteristic equation of $N-X^{\rm R}$ is 
given by
\begin{eqnarray*}
\lambda\left(\lambda-\frac{\left|\braket{\phi_{1}}{\phi_{2}}\right|}{\sqrt{\eta_{1}\eta_{2}}}\right)=0, 
\end{eqnarray*}
which clearly shows that the two eigenvalues are non-negative. 
Therefore, $x^{\rm R}_{i}$ is the solution of the relaxed two-state problem.

Thus, if $x^{\rm R}_{i}$ are all non-negative, they give the optimal solution of 
the two-state problem $x_{i}$ as 
\begin{align}
x_{1}&=1-\sqrt{\frac{\eta_{2}}{\eta_{1}}}\left|\braket{\phi_{1}}{\phi_{2}}\right|, \\
x_{2}&=1-\sqrt{\frac{\eta_{1}}{\eta_{2}}}\left|\braket{\phi_{1}}{\phi_{2}}\right|, 
\end{align}
and the maximum success probability $p_{\max}$ is given by $d_{\min}$,  
\begin{eqnarray}
p_{\max}=1-2\sqrt{\eta_{1}\eta_{2}}\left|\braket{\phi_{1}}{\phi_{2}}\right|. 
\end{eqnarray}

If one of $x^{\rm R}_{i}$ is negative, however, 
omitting one of the states to be discriminated is optimal. 
Let us first assume $x^{\rm R}_{2} < 0$, which happens when the occurrence probabilities 
and the inner product satisfy the following condition:  
\begin{align}
\sqrt{\frac{\eta_{2}}{\eta_{1}}} < \left|\braket{\phi_{1}}{\phi_{2}}\right|. 
\end{align}
We then need to search for the optimal solution with $x_{2}=0$. 
According to Eqs. (\ref{eq:phi-prime}) and (\ref{eq:eta-prime}), the two-state problem with the additional constraint $x_{2}=0$ is 
equivalent to the one-state problem for the following state $\ket{\phi'_{1}}$ 
with occurrence probability $\eta'_{1}$:
\begin{align}
\ket{\phi'_{1}}=\frac{Q\ket{\phi_{1}}}{\sqrt{\bra{\phi_{1}}Q\ket{\phi_{1}}}}, \hspace{1ex}\eta'_{1}=\frac{\eta_{1}\bra{\phi_{1}}Q\ket{\phi_{1}}}{\eta_{1}\bra{\phi_{1}}Q\ket{\phi_{1}}}=1, 
\end{align}
where $Q$ is the projector defined by $Q\equiv\bold{1}-\ket{\phi_{2}}\bra{\phi_{2}}$. 
The maximum success probabilities and the variables $x_{i}$ and $x'_{i}$ of the 
two problems are related as follows:
\begin{align*}
p^{(2)}_{\max}[\{\eta_{i}, \ket{\phi_{i}}\}^{2}_{i=1}] 
    &= \eta_{1}\bra{\phi_{1}}Q\ket{\phi_{1}}
          p^{(1)}_{\max}[\{\eta'_{1}, \ket{\phi'_{1}}\}], \\
x_{1} &= \bra{\phi_{1}}Q\ket{\phi_{1}}x'_{1}, 
\end{align*}
where $x'_{1}$ is the solution of the one-state problem. 

The one-state problem is trivial: 
$p^{(1)}_{\max}[\{\eta'_{1}, \ket{\phi'_{1}}\}]=1$ and $x'_{1}=1$, 
since there is only one state to be discriminated. Thus, the maximum success probability $p_{\max}$ is obtained to be 
\begin{eqnarray}
p_{\max}=\eta_{1}\left(1-\left|\braket{\phi_{1}}{\phi_{2}}\right|^{2}\right) 
\end{eqnarray} 
and $x_{i}\hspace{0.5ex}(i=1,\hspace{0.5ex}2)$ is given by
\begin{subequations} 
\begin{align}
x_{1}&=1-\left|\braket{\phi_{1}}{\phi_{2}}\right|^{2}, \\
x_{2}&=0. 
\end{align} 
\end{subequations}

We assumed that $x^{\rm R}_{2} < 0$. For the case of $x^{\rm R}_{1} < 0$, it is clear that this problem becomes a similar two-state problem with the additional constraint $x_{1}=0$.

The maximum success probability and the optimal solution $x_i$ are summarized in the 
following way:
\begin{align}
&\hspace{-10ex}(i)\hspace{2ex}\sqrt{\frac{\eta_{2}}{\eta_{1}}} < \left|\braket{\phi_{1}}{\phi_{2}}\right|\nonumber\\
&p_{\max}=\eta_{1}\left(1-\left|\braket{\phi_{1}}{\phi_{2}}\right|^{2}\right),
\end{align}
\begin{subequations}
\begin{align}
&x_{1}=1-\left|\braket{\phi_{1}}{\phi_{2}}\right|^{2},\\
&x_{2}=0.
\end{align}
\end{subequations}
\begin{align}
&\hspace{-10ex}(ii)\hspace{2ex}\left|\braket{\phi_{1}}{\phi_{2}}\right| \leq \sqrt{\frac{\eta_{2}}{\eta_{1}}} \leq \frac{1}{\left|\braket{\phi_{1}}{\phi_{2}}\right|}\nonumber\\
&p_{\max}=1-2\sqrt{\eta_{1}\eta_{2}}\left|\braket{\phi_{1}}{\phi_{2}}\right|,
\end{align}
\begin{subequations}
\begin{align}
&x_{1}=1-\sqrt{\frac{\eta_{2}}{\eta_{1}}}\left|\braket{\phi_{1}}{\phi_{2}}\right|,\\
&x_{2}=1-\sqrt{\frac{\eta_{1}}{\eta_{2}}}\left|\braket{\phi_{1}}{\phi_{2}}\right|.
\end{align}
\end{subequations}
\begin{align}
&\hspace{-10ex}(iii)\hspace{2ex}\frac{1}{\left|\braket{\phi_{1}}{\phi_{2}}\right|} < \sqrt{\frac{\eta_{2}}{\eta_{1}}}\nonumber\\
&p_{\max}=\eta_{2}\left(1-\left|\braket{\phi_{1}}{\phi_{2}}\right|^{2}\right),
\end{align}
\begin{subequations}
\begin{align}
&x_{1}=0,\\
&x_{2}=1-\left|\braket{\phi_{1}}{\phi_{2}}\right|^{2}. 
\end{align}
\end{subequations}

This reproduces the well-known result of the unambiguous discrimination problem between 
two pure states. These results will be used in the subsequent section.

\subsection{Three pure states} 
We consider the unambiguous discrimination problem between three pure states, $\ket{\phi_{1}}$, $\ket{\phi_{2}}$, and $\ket{\phi_{3}}$, with occurrence probabilities $\eta_{1}$, 
$\eta_{2}$, and $\eta_{3}$, respectively. 

The inner products between the states depend on phases of the states. 
However, $\Gamma$ defined by
\begin{eqnarray}
\Gamma \equiv 
\braket{\phi_{1}}{\phi_{2}}\braket{\phi_{2}}{\phi_{3}}\braket{\phi_{3}}{\phi_{1}},  
\end{eqnarray}
is invariant under any phase change of each state.  A little thought shows that the phases 
of the three mutual inner products can be chosen to be the same by appropriate phase 
changes of the states. 

Here, we assume that $\Gamma$ is real, implying that the three mutual inner products 
can be chosen to be real. We separately consider two cases of 
$\Gamma \leq 0$ and $\Gamma > 0$. The case of complex $\Gamma$ will be discussed 
in the next section.  


\subsubsection{Case of $\Gamma \leq 0$} \label{sec:Gamma_negative}
In the case of $\Gamma \leq 0$, the phases of the mutual inner products can be chosen 
to be real and negative.  As we will see, we can proceed along the same lines as in 
the two-state problem.
The $d$ is given by
\begin{align}
d =& \tr{YN} \nonumber\\ 
  =& 1-\left(\braket{y_{1}}{y_{2}}+\braket{y_{2}}{y_{1}}\right)
               \left|\braket{\phi_{1}}{\phi_{2}}\right| \nonumber\\
   &\ \,-\left(\braket{y_{2}}{y_{3}}+\braket{y_{3}}{y_{2}}\right)
               \left|\braket{\phi_{2}}{\phi_{3}}\right| \nonumber\\
   &\ \,-\left(\braket{y_{3}}{y_{1}}+\braket{y_{1}}{y_{3}}\right)
               \left|\braket{\phi_{3}}{\phi_{1}}\right|.  
                       \label{eq:minimizing_d_of_three_state}
\end{align} 
Using the triangle inequality and Schwarz's inequality as in the two-state case, 
we immediately obtain the minimum of $d$ to be
\begin{align}
d_{\min} = 
   1 & -2\sqrt{\eta_{1}\eta_{2}}\left|\braket{\phi_{1}}{\phi_{2}}\right|
       -2\sqrt{\eta_{2}\eta_{3}}\left|\braket{\phi_{2}}{\phi_{3}}\right| \nonumber \\
     & -2\sqrt{\eta_{3}\eta_{1}}\left|\braket{\phi_{3}}{\phi_{1}}\right|,
\end{align}
and the vectors $\ket{y_{i}}$ that attain this lower bound are the following 
one-dimensional vectors: 
\begin{align*}
\ket{y_{1}}= \sqrt{\eta_{1}},\hspace{1ex}\ket{y_{2}}= \sqrt{\eta_{2}},\hspace{1ex}\ket{y_{3}}= \sqrt{\eta_{3}}.
\end{align*}
Thus, $Y$ can be expressed as 
\begin{eqnarray}
Y = \mbold{\beta}\mbold{\beta}^{\dagger}, \label{eq:Y_of_three_state_of_negative}
\end{eqnarray}
where $\mbold{\beta}$ is defined by
\begin{eqnarray*}
\mbold{\beta}=\left(
\begin{array}{c}
\sqrt{\eta_{1}}\\
\sqrt{\eta_{2}}\\
\sqrt{\eta_{3}}\\
\end{array}
\right).
\end{eqnarray*}

The attainability condition (\ref{eq:attainability}) again takes the form, 
\begin{eqnarray}
\left(N-X^{\rm R}\right)\mbold{\beta}=0 .\label{eq:attain1}
\end{eqnarray}
since $\mbold{\beta}^{\dagger}\mbold{\beta}=1$.  
In the matrix representation, Eq. (\ref{eq:attain1}) is written as
\begin{align*}
&
\left(
\begin{array}{ccc}
1-x^{\rm R}_{1} & -\left|\braket{\phi_{1}}{\phi_{2}}\right| & -\left|\braket{\phi_{3}}{\phi_{1}}\right|\\
-\left|\braket{\phi_{1}}{\phi_{2}}\right| & 1-x^{\rm R}_{2} & -\left|\braket{\phi_{2}}{\phi_{3}}\right|\\
-\left|\braket{\phi_{3}}{\phi_{1}}\right| & -\left|\braket{\phi_{2}}{\phi_{3}}\right| & 1-x^{\rm R}_{3}\\
\end{array}
\right)\left(
\begin{array}{c}
\sqrt{\eta_{1}}\\
\sqrt{\eta_{2}}\\
\sqrt{\eta_{3}}\\
\end{array}
\right) \\
& =0 .
\end{align*}

Solving this equation, we obtain $x^{\rm R}_{1}$, $x^{\rm R}_{2}$, and $x^{\rm R}_{3}$ 
to be 
\begin{align*}
x^{\rm R}_{1}=1-\frac{1}{\sqrt{\eta_{1}}}\left(\sqrt{\eta_{2}}\left|\braket{\phi_{1}}{\phi_{2}}\right|+\sqrt{\eta_{3}}\left|\braket{\phi_{3}}{\phi_{1}}\right|\right), \\
x^{\rm R}_{2}=1-\frac{1}{\sqrt{\eta_{2}}}\left(\sqrt{\eta_{1}}\left|\braket{\phi_{1}}{\phi_{2}}\right|+\sqrt{\eta_{3}}\left|\braket{\phi_{2}}{\phi_{3}}\right|\right), \\
x^{\rm R}_{3}=1-\frac{1}{\sqrt{\eta_{3}}}\left(\sqrt{\eta_{1}}\left|\braket{\phi_{3}}{\phi_{1}}\right|+\sqrt{\eta_{2}}\left|\braket{\phi_{2}}{\phi_{3}}\right|\right). 
\end{align*}

Now we check if $x^{\rm R}_{i}$ satisfy the condition (\ref{eq:relaxed-condition}), 
that is, the positivity of $N-X^{\rm R}$. 
The attainability condition (\ref{eq:attain1}) shows that one of the 
eigenvalues of $N-X^{\rm R}$ is 0. The remaining two eigenvalues are determined by the 
following quadratic equation:
\begin{align*}
  \lambda^2 - a \lambda + b = 0, 
\end{align*}
where the coefficients $a$ and $b$ are given by 
\begin{align*}
  a =& \frac{\eta_{1}+\eta_{2}}{\sqrt{\eta_{1}\eta_{2}}}
          \left|\braket{\phi_{1}}{\phi_{2}}\right|
      +\frac{\eta_{2}+\eta_{3}}{\sqrt{\eta_{2}\eta_{3}}}
          \left|\braket{\phi_{2}}{\phi_{3}}\right| \\
     &+\frac{\eta_{3}+\eta_{1}}{\sqrt{\eta_{3}\eta_{1}}}
          \left|\braket{\phi_{3}}{\phi_{1}}\right|, \\
  b =& \frac{\left|\braket{\phi_{2}}{\phi_{3}}
                   \braket{\phi_{3}}{\phi_{1}}\right|}{\sqrt{\eta_{1}\eta_{2}}}
      +\frac{\left|\braket{\phi_{1}}{\phi_{2}}
                   \braket{\phi_{3}}{\phi_{1}}\right|}{\sqrt{\eta_{2}\eta_{3}}} \\
     &+\frac{\left|\braket{\phi_{1}}{\phi_{2}}
                   \braket{\phi_{2}}{\phi_{3}}\right|}{\sqrt{\eta_{3}\eta_{1}}}.         
\end{align*} 
The two roots, the eigenvalues of the real symmetric matrix, must be real, and the quadratic formula shows that they are 
both non-negative since $a \geq 0$ and $b \geq 0$.
Thus, the positivity of $N-X^{\rm R}$ is satisfied, and $x^{\rm R}_{i}$ is therefore 
the solution of the relaxed problem. 

If $x^{\rm R}_{i}$ are all non-negative, $x^{\rm R}_{i}$ is also the optimal solution 
of the original three-state problem, 
\begin{subequations}
\begin{align}
x_{1}=1-\frac{1}{\sqrt{\eta_{1}}}\left(\sqrt{\eta_{2}}\left|\braket{\phi_{1}}{\phi_{2}}\right|+\sqrt{\eta_{3}}\left|\braket{\phi_{3}}{\phi_{1}}\right|\right),\\
x_{2}=1-\frac{1}{\sqrt{\eta_{2}}}\left(\sqrt{\eta_{1}}\left|\braket{\phi_{1}}{\phi_{2}}\right|+\sqrt{\eta_{3}}\left|\braket{\phi_{2}}{\phi_{3}}\right|\right),\\
x_{3}=1-\frac{1}{\sqrt{\eta_{3}}}\left(\sqrt{\eta_{1}}\left|\braket{\phi_{3}}{\phi_{1}}\right|+\sqrt{\eta_{2}}\left|\braket{\phi_{2}}{\phi_{3}}\right|\right), 
\end{align}
\end{subequations}
and the optimal success probability $p_{\max}$ is given by $d_{\min}$ as follows:
\begin{align}
p_{\max}=1 & -2\sqrt{\eta_{1}\eta_{2}}\left|\braket{\phi_{1}}{\phi_{2}}\right|
             -2\sqrt{\eta_{2}\eta_{3}}\left|\braket{\phi_{2}}{\phi_{3}}\right| 
                          \nonumber \\
           & -2\sqrt{\eta_{3}\eta_{1}}\left|\braket{\phi_{3}}{\phi_{1}}\right|. 
\end{align}

If some of $x^{\rm R}_{i}$ are negative, the problem can be 
reduced to a two-state problem. The two-state ensemble 
$\{\eta'_{i}, \ket{\phi'_{i}}\}^{2}_{i=1}$ to be considered is determined by general 
formulas of Eqs.~(\ref{eq:phi-prime}) and (\ref{eq:eta-prime}). 
The maximum success probability and the solution $x_i$ of the original problem can be 
obtained by Eqs.~(\ref{eq:p-max-reduction}) and (\ref{eq:x-i-reduction}) in terms of 
the well-known solutions of two-state problems. 

\subsubsection{Case of $\Gamma > 0$} \label{sec:Gamma_positive}
In the case of $\Gamma > 0$, the phases of the mutual inner products can be chosen to be real 
and positive. 
We introduce three positive real numbers $\alpha_{1}$, $\alpha_{2}$, and $\alpha_{3}$ as 
\begin{align}
\alpha_{1} &\equiv \sqrt{\frac{\braket{\phi_{1}}{\phi_{2}}\braket{\phi_{3}}{\phi_{1}}}{\braket{\phi_{2}}{\phi_{3}}}},\\
\alpha_{2} &\equiv \sqrt{\frac{\braket{\phi_{1}}{\phi_{2}}\braket{\phi_{2}}{\phi_{3}}}{\braket{\phi_{3}}{\phi_{1}}}},\\
\alpha_{3} &\equiv \sqrt{\frac{\braket{\phi_{2}}{\phi_{3}}\braket{\phi_{3}}{\phi_{1}}}{\braket{\phi_{1}}{\phi_{2}}}}.
\end{align}
By using $\alpha_{i}$, the inner products can be expressed as follows:
\begin{align*}
\braket{\phi_{1}}{\phi_{2}}=\alpha_{1}\alpha_{2},\hspace{1ex}\braket{\phi_{2}}{\phi_{3}}=\alpha_{2}\alpha_{3},\hspace{1ex}\braket{\phi_{3}}{\phi_{1}}=\alpha_{3}\alpha_{1}.
\end{align*}
The $d$ can then be rewritten as 
\begin{align}
d &= \sum_{i,j=1}^3 Y_{ij}N_{ji} \nonumber \\
  &= 1+\sum^{3}_{i \neq j}\alpha_{i}\alpha_{j}\braket{y_{i}}{y_{j}}\nonumber\\
  &= 1-\sum^{3}_{i=1}\eta_{i}\alpha_{i}^{2}
         +\left|\sum^{3}_{i=1}\alpha_{i}\ket{y_{i}}\right|^{2}. \label{eq:min_d2}
\end{align}

Now, we must minimize $d$. This means that we must minimize $\left|\sum^{3}_{i=1}\alpha_{i}\ket{y_{i}}\right|$,  
which is the norm of the sum of three vectors $\alpha_{1}\ket{y_{1}}$, 
$\alpha_{2}\ket{y_{2}}$, and $\alpha_{3}\ket{y_{3}}$. Here $\ket{y_{i}}$ is any vector with the norm $\sqrt{\eta_{i}}$. The minimum value of $\left|\sum^{3}_{i=1}\alpha_{i}\ket{y_{i}}\right|$ is determined depending on 
whether the three lengths $\alpha_i\sqrt{\eta_i}$ satisfy the following 
set of relations:  
\begin{subequations}
\label{eq:triangle_relation}
\begin{align}
\alpha_{1}\sqrt{\eta_{1}} \leq \alpha_{2}\sqrt{\eta_{2}}+\alpha_{3}\sqrt{\eta_{3}},\label{eq:triangle1}\\
\alpha_{2}\sqrt{\eta_{2}} \leq \alpha_{1}\sqrt{\eta_{1}}+\alpha_{3}\sqrt{\eta_{3}},\\
\alpha_{3}\sqrt{\eta_{3}} \leq \alpha_{1}\sqrt{\eta_{1}}+\alpha_{2}\sqrt{\eta_{2}}.
\end{align}
\end{subequations}
These relations are the condition for the three vectors $\alpha_{i}\ket{y_{i}}$ to 
form a triangle, and we call it the triangle condition hereafter.  

First, we assume that the triangle condition (\ref{eq:triangle_relation})
is satisfied. 
The minimum value of $\left|\sum^{3}_{i=1}\alpha_{i}\ket{y_{i}}\right|$ is 0, achieved by one-dimensional complex vector $\ket{y_{i}}=y_{i}$ with $\left|y_{i}\right|=\sqrt{\eta_{i}}$ such that $\alpha_{1}y_{1}+\alpha_{2}y_{2}+\alpha_{3}y_{3}=0$. 
The minimum value of $d$ is then given by
\begin{align}
d_{\min}=1-\sum^{3}_{i=1}\eta_{i}\alpha_{i}^{2}. 
\end{align}
Writing
\begin{align*}
\mbold{y}=\left(
\begin{array}{c}
y_{1}\\
y_{2}\\
y_{3}
\end{array}
\right),
\end{align*}
we have $Y=\mbold{y}^{\ast}\mbold{y}^{T}$.
Thus, Eq.~(\ref{eq:attainability}) is rewritten as
\begin{eqnarray*}
\left(N-X^\rmR \right)\mbold{y}^{\ast}=0,  
\end{eqnarray*}
from which we obtain the following three equations for $x^{\rm R}_{1}$, $x^{\rm R}_{2}$, 
and $x^{\rm R}_{3}$:
\begin{align}
\left(1-x^{\rm R}_{1}\right)y_{1}^{\ast}+\alpha_{1}\alpha_{2}y_{2}^{\ast}+\alpha_{1}\alpha_{3}y_{3}^{\ast}=0,\\
\alpha_{2}\alpha_{1}y_{1}^{\ast}+\left(1-x^{\rm R}_{2}\right)y_{2}^{\ast}+\alpha_{2}\alpha_{3}y_{3}^{\ast}=0,\\
\alpha_{3}\alpha_{1}y_{1}^{\ast}+\alpha_{3}\alpha_{2}y_{2}^{\ast}+\left(1-x^{\rm R}_{3}\right)y_{3}^{\ast}=0. 
\end{align}
Now, remember that $y_i$ are chosen in such a way that  $\alpha_{1}y_{1}+\alpha_{2}y_{2}+\alpha_{3}y_{3}=0$. 
Thus, the first equation is given by 
\begin{align*}
\left(1-x^{\rm R}_{1}-\alpha_{1}^{2}\right)y_{1}^{\ast}=0. 
\end{align*}
Since $y_{1}^{\ast} \neq 0$, we obtain $x^{\rm R}_{1}=1-\alpha_{1}^{2}$. 
Similarly, we obtain $x^{\rm R}_{2}$ and $x^{\rm R}_{3}$.   
\begin{align*}
x^{\rm R}_{1}=1-\alpha_{1}^{2},\\
x^{\rm R}_{2}=1-\alpha_{2}^{2},\\
x^{\rm R}_{3}=1-\alpha_{3}^{2}. 
\end{align*}

We can easily see the positivity of $N-X^\rmR$, since $N-X^\rmR$ 
is expressed as $\mbold{\alpha}\mbold{\alpha}^\dagger$, 
where $\mbold{\alpha}$ is defined by
\begin{eqnarray*}
\mbold{\alpha}=\left(
\begin{array}{c}
\alpha_{1}\\
\alpha_{2}\\
\alpha_{3}\\
\end{array}
\right).
\end{eqnarray*}
Thus, $x^{\rm R}_{i}$ we obtained is the optimal solution of the relaxed three-state 
problem.  

If $x^{\rm R}_{i}$ are all non-negative, 
they are the optimal solution $x_{i}$ of the original three state problem, 
\begin{subequations}
\label{eq:the_optimal_solution_of_triangle_relation_satisfy}
\begin{align}
x_{1}=1-\alpha_{1}^{2},\\
x_{2}=1-\alpha_{2}^{2},\\
x_{3}=1-\alpha_{3}^{2}.
\end{align}
\end{subequations}
and the success probability $p_{\max}$ is given by $d_{\min}$ as follows: 
\begin{eqnarray}
p_{\max}=1-\left(\eta_{1}\alpha_{1}^{2}+\eta_{2}\alpha_{2}^{2}+\eta_{3}\alpha_{3}^{2}\right). 
\end{eqnarray}
If some of $x^{\rm R}_{i}$ turn out to be negative, we can solve the problem by 
reducing it to the two-state problem as described before. 


Next, we consider the case in which the triangle relation (\ref{eq:triangle_relation}) 
is not satisfied. Comparing the norms of $\alpha_{1}\ket{y_{1}}$, $\alpha_{2}\ket{y_{2}}$,  and $\alpha_{3}\ket{y_{3}}$, we realize that the norm 
$\left|\sum^{3}_{i=1}\alpha_{i}\ket{y_{i}}\right|$ is minimized by  
subtracting the two smaller norms from the largest one. Suppose that $\alpha_{1}\sqrt{\eta_{1}} \geq \alpha_{2}\sqrt{\eta_{2}},\ \alpha_{3}\sqrt{\eta_{3}}$. Then, the minimum value $d_{\min}$ is given by
\begin{align}
& d_{\min}=1-\sum^{3}_{i=1}\eta_{i}\alpha_{i}^{2}
   +\left(\alpha_{1}\sqrt{\eta_{1}}-\alpha_{2}\sqrt{\eta_{2}}
            -\alpha_{3}\sqrt{\eta_{3}}\right)^{2}\nonumber\\
& =1-2\sqrt{\eta_{1}\eta_{2}}\alpha_{1}\alpha_{2}
    +2\sqrt{\eta_{2}\eta_{3}}\alpha_{2}\alpha_{3}
    -2\sqrt{\eta_{3}\eta_{1}}\alpha_{3}\alpha_{1}, 
\end{align}
and the vectors $\ket{y_{i}}$ that attain this minimum value are the following 
one-dimensional vectors: 
\begin{align*}
\ket{y_{1}}= \sqrt{\eta_{1}},\hspace{1ex}\ket{y_{2}}=-\sqrt{\eta_{2}},\hspace{1ex}\ket{y_{3}}=-\sqrt{\eta_{3}}.
\end{align*}
Thus, $Y$ is given by $\mbold{\beta}\mbold{\beta}^{\dagger}$ 
with $\mbold{\beta}$ defined to be 
\begin{eqnarray*}
\mbold{\beta}=\left(
\begin{array}{c}
\sqrt{\eta_{1}}\\
-\sqrt{\eta_{2}}\\
-\sqrt{\eta_{3}}\\
\end{array}
\right) .
\end{eqnarray*}

The rest of the argument proceeds in a quite similar way to that in the case 
of $\Gamma \le 0$. 
Since $Y=\mbold{\beta}\mbold{\beta}^{\dagger}$, the attainability condition Eq.~(\ref{eq:attainability}) can be rewritten as
\begin{eqnarray}
\left(N-X^{\rm R}\right)\mbold{\beta}=0, \label{eq:attain2}
\end{eqnarray}
with its matrix representation given by 
\begin{eqnarray*}
\left(
\begin{array}{ccc}
1-x^{\rm R}_{1} & \alpha_{1}\alpha_{2} & \alpha_{1}\alpha_{3}\\
\alpha_{2}\alpha_{1} & 1-x^{\rm R}_{2} & \alpha_{2}\alpha_{3}\\
\alpha_{3}\alpha_{1} & \alpha_{3}\alpha_{2} & 1-x^{\rm R}_{3}\\
\end{array}
\right)\left(
\begin{array}{c}
\sqrt{\eta_{1}}\\
-\sqrt{\eta_{2}}\\
-\sqrt{\eta_{3}}\\
\end{array}
\right)=0.
\end{eqnarray*}
Solving this equation, we obtain $x^{\rm R}_{i}$ as follows:
\begin{align*}
x^{\rm R}_{1}&=1-\frac{\alpha_{1}}{\sqrt{\eta_{1}}}\left(\alpha_{2}\sqrt{\eta_{2}}+\alpha_{3}\sqrt{\eta_{3}}\right),\\
x^{\rm R}_{2}&=1-\frac{\alpha_{2}}{\sqrt{\eta_{2}}}\left(\alpha_{1}\sqrt{\eta_{1}}-\alpha_{3}\sqrt{\eta_{3}}\right),\\
x^{\rm R}_{3}&=1-\frac{\alpha_{3}}{\sqrt{\eta_{3}}}\left(\alpha_{1}\sqrt{\eta_{1}}-\alpha_{2}\sqrt{\eta_{2}}\right).
\end{align*}

The positivity of $N-X^{\rm R}$ can be verified in the following way: 
One of the eigenvalues of $N-X^{\rm R}$ is 0, which can be seen from 
Eq.~(\ref{eq:attain2}). The remaining two eigenvalues are given by the roots of  
the following quadratic equation:
\begin{align*}
  \lambda^2-a\lambda+b =0, 
\end{align*}
where
\begin{align*}
 a =& \frac{\alpha_{1}}{\sqrt{\eta_{1}}}
         \left(\alpha_{2}\sqrt{\eta_{2}}+\alpha_{3}\sqrt{\eta_{3}}\right)
     +\frac{\alpha_{2}}{\sqrt{\eta_{2}}}
         \left(\alpha_{1}\sqrt{\eta_{1}}-\alpha_{3}\sqrt{\eta_{3}}\right) 
                 \\
    &+\frac{\alpha_{3}}{\sqrt{\eta_{3}}}
         \left(\alpha_{1}\sqrt{\eta_{1}}-\alpha_{2}\sqrt{\eta_{2}}\right),
                 \\
 b =& \frac{\alpha_{1}\alpha_{2}\alpha_{3}}{\sqrt{\eta_{1}\eta_{2}\eta_{3}}}
       \left(\alpha_{1}\sqrt{\eta_{1}}-\alpha_{2}\sqrt{\eta_{2}}
            -\alpha_{3}\sqrt{\eta_{3}}\right).
\end{align*} 
Remember that we assumed that 
$\alpha_{1}\sqrt{\eta_{1}} \geq \alpha_{2}\sqrt{\eta_{2}},\ \alpha_{3}\sqrt{\eta_{3}}$, 
which implies that the relation (\ref{eq:triangle1}) is violated. Thus, we have $a \ge 0$ and $b \ge 0$.  
Therefore, the remaining two eigenvalues are also non-negative. 
Thus, $x_i^\rmR$ are the solution of the relaxed two-state problem. 

If $x^{\rm R}_{i}$ are all non-negative, $x_i^\rmR$ is the optimal solution $x_i$ 
of the original three-state problem, 
\begin{subequations} 
\begin{align}
x_{1}&=1-\frac{1}{\sqrt{\eta_{1}}}\left(\sqrt{\eta_{2}}\left|\braket{\phi_{1}}{\phi_{2}}\right|+\sqrt{\eta_{3}}\left|\braket{\phi_{3}}{\phi_{1}}\right|\right),\\
x_{2}&=1-\frac{1}{\sqrt{\eta_{2}}}\left(\sqrt{\eta_{1}}\left|\braket{\phi_{1}}{\phi_{2}}\right|-\sqrt{\eta_{3}}\left|\braket{\phi_{2}}{\phi_{3}}\right|\right),\\
x_{3}&=1-\frac{1}{\sqrt{\eta_{3}}}\left(\sqrt{\eta_{1}}\left|\braket{\phi_{3}}{\phi_{1}}\right|-\sqrt{\eta_{2}}\left|\braket{\phi_{2}}{\phi_{3}}\right|\right), 
\end{align}
\end{subequations}
and the maximum success probability $p_{\max}$ is given by $d_{\min}$, 
\begin{align}
 p_{\max}=1 & -2\sqrt{\eta_{1}\eta_{2}}\left|\braket{\phi_{1}}{\phi_{2}}\right|
              +2\sqrt{\eta_{2}\eta_{3}}\left|\braket{\phi_{2}}{\phi_{3}}\right| \nonumber \\
            & -2\sqrt{\eta_{3}\eta_{1}}\left|\braket{\phi_{3}}{\phi_{1}}\right|. 
\end{align}
Note that apparent asymmetry of the signs of terms in the above results  
is due to our assumption that 
$\alpha_{1}\sqrt{\eta_{1}} \geq \alpha_{2}\sqrt{\eta_{2}},\ \alpha_{3}\sqrt{\eta_{3}}$. 

If some of $x_i^\rmR$ are negative, we can still have analytical results by considering 
the appropriate two-state problem in the same way as in other cases.

\subsection{Examples}
We consider the set of the following three-state vectors: 
\begin{align*}
&
\ket{\phi_{1}}=\left(
\begin{array}{c}
1\\
0\\
0
\end{array}
\right),\hspace{1ex}\ket{\phi_{2}}=\left(
\begin{array}{c}
\cos\varphi_{2}\\
\sin\varphi_{2}\\
0
\end{array}
\right), \\
&
\ket{\phi_{3}}=\left(
\begin{array}{c}
\cos\varphi_{3}\sin\theta_{3}\\
\sin\varphi_{3}\sin\theta_{3}\\
\cos\theta_{3}
\end{array}
\right).
\end{align*}

\begin{figure}
\begin{center}
\includegraphics[width=8cm]{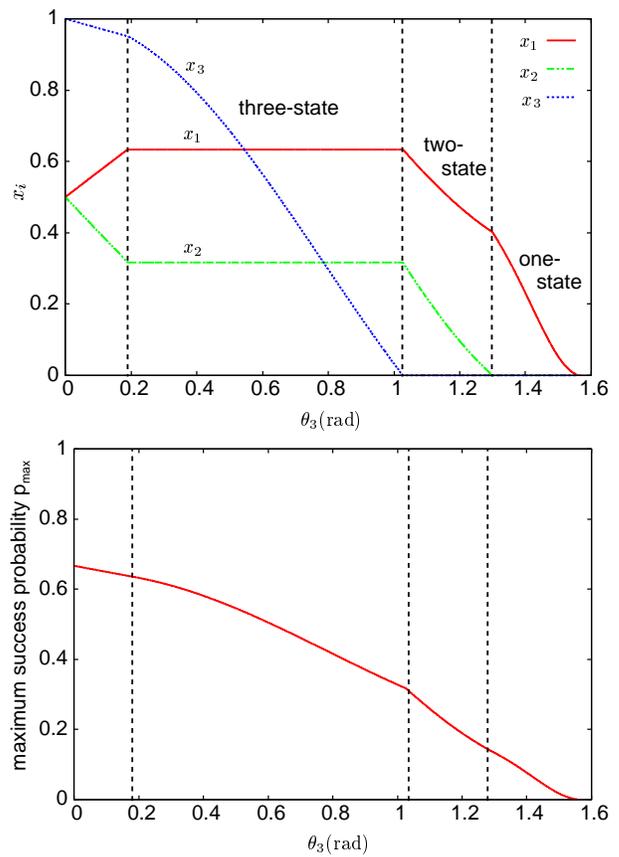}
\caption{(Color online) Example 1. $x_{i}$ (upper part) and the maximum success 
probability (lower part) vs. $\theta_{3}$. The angles $\varphi_{2}$ and $\varphi_{3}$ 
are fixed to be $\pi/3$ and $\pi/4$, respectively.}
\label{fig:theta}
\end{center}
\end{figure}

In the first example, the equal occurrence probabilities are assumed, and  
$\varphi_{2}$ and $\varphi_{3}$ are fixed to be $\pi/3$ and 
$\pi/4$, respectively. 
Figure~\ref{fig:theta} displays the optimal solution $x_i$ and the maximum 
success probability as a function of $\theta_{3}$. As $\theta_{3}$ varies from 0 to 
$\pi/2$, the type of solutions changes. When $\theta_{3}$ is small, all the three states 
have finite identification probabilities $x_i$. As $\theta_{3}$ increases, the number of 
the states with nonvanishing $x_i$ decreases. This reflects the fact that 
the state $\ket{\phi_{3}}$ approaches the state $\ket{\phi_{2}}$ and it becomes 
difficult to discriminate between these two states.
When $\theta_{3}$ is in the range from 0.18(rad) to 1.01(rad), $x_{1}$ and $x_{2}$ are constant. 
That is when the three lengths $\alpha_{i}\sqrt{\eta_{i}}$ satisfy the triangle 
condition (\ref{eq:triangle_relation}) in the case of $\Gamma > 0$. 
In such a case, $x_{i}$ is expressed only by the mutual inner products as in 
Eq.(\ref{eq:the_optimal_solution_of_triangle_relation_satisfy}). 
\begin{align*}
x_{1}&=1-\frac{\cos\varphi_{2}\cos\varphi_{3}}{\cos\left(\varphi_{2}-\varphi_{3}\right)},\\
x_{2}&=1-\frac{\cos\varphi_{2}\cos\left(\varphi_{2}-\varphi_{3}\right)}{\cos\varphi_{3}},\\
x_{3}&=1-\frac{\cos\varphi_{3}\cos\left(\varphi_{2}-\varphi_{3}\right)}{\cos\varphi_{2}}\sin^{2}\theta_{3}.
\end{align*}
It is clear that $x_{1}$ and $x_{2}$ are independent of $\theta_{3}$. Thus, $x_{1}$ and $x_{2}$ are constant in this range.  
 
\begin{figure}
\begin{center}
\includegraphics[width=8cm]{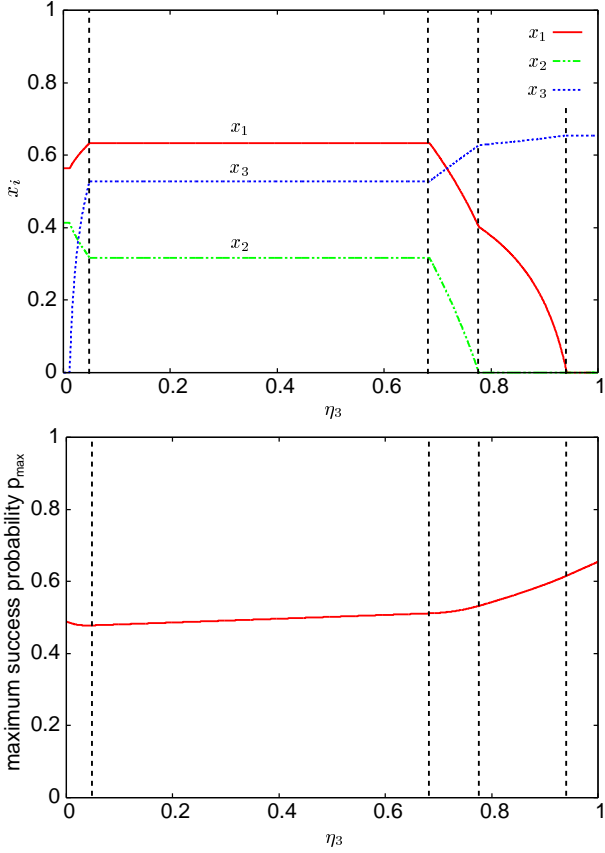}
\caption{(Color online) Example 2. $x_{i}$ (upper part) and the maximum success 
probability (lower part) vs. $\eta_{3}$. $\eta_1=\eta_2$ is assumed. The three states 
are fixed by setting $\varphi_{2}=\pi/3$, $\varphi_{3}=\pi/4$, and $\theta_{3}=\pi/5$. }
\label{fig:eta3}
\end{center}
\end{figure}

In the second example, the occurrence probabilities are varied while the three states 
are fixed by setting $\varphi_{2}=\pi/3$, $\varphi_{3}=\pi/4$, and $\theta_{3}=\pi/5$. 
Assuming $\eta_1=\eta_2$, we vary $\eta_3$ from 0 to 1. 
Figure~\ref{fig:eta3} displays the optimal solution $x_i$ and the maximum 
success probability as a function of $\eta_{3}$. 
For a sufficiently small $\eta_{3}$, the optimal solution gives $x_{3}=0$. 
As $\eta_3$ approaches 1, only the state $\ket{\phi_3}$ survives as expected. 
However,  the maximum success probability does not approach 1. 
This is because we assumed $\eta_i \ne 0$ and imposed the no-error condition 
(\ref{eq:no-error-condition}) for all states even if some of $\eta_i$ are very small. 
One interesting thing is that if $\Gamma > 0$ and the triangle condition (\ref{eq:triangle_relation}) is satisfied, 
Eq.~(\ref{eq:the_optimal_solution_of_triangle_relation_satisfy}) shows that $x_i$ 
does not depend on $\eta_i$: The optimal measurement is independent of the occurrence probabilities. This is the reason why all the $x_{i}$ are constant when $\eta_{3}$ 
ranges from 0.05 to 0.7.  

Before concluding the section, we mention that the problem of unambiguous discrimination 
can be numerically solved since the problem is formulated as semidefinite programming. 
We performed a numerical analysis for 
the above examples using the semidefinite programming package SDPA \cite{Yamashita03}, and we compared the two results just to be sure. 
We verified that the numerical results reproduced our analytical ones 
within numerical errors. 

\section{Discussion}
\label{sec_Discussion} 

In the preceding section, we have obtained complete analytical solutions for 
unambiguous discrimination of three pure states with real mutual inner products 
and general occurrence probabilities. Here, we examine what difficulties arise 
when the inner products are complex. Though our method is not applicable to 
general complex cases, we will show that there is one instance for which 
an analytical result can be obtained.  
 
As in the preceding section, we define $\Gamma$ in the following way:
\begin{align}
  \Gamma = \braket{\phi_1}{\phi_2}\braket{\phi_2}{\phi_3}\braket{\phi_3}{\phi_1}
         =  |\Gamma| e^{3i\theta},  
\end{align} 
where $0 \le \theta < 2\pi/3$. We can evidently choose phases of the three states  
so that the phases of inner products $\braket{\phi_1}{\phi_2}$, 
$\braket{\phi_2}{\phi_3}$, and $\braket{\phi_3}{\phi_1}$ are equal to the same 
value $\theta$. Then, we can express the three inner products in the 
following way: 
\begin{align}
   \braket{\phi_1}{\phi_2} &= \alpha_1 \alpha_2 e^{i\theta}, \\
   \braket{\phi_2}{\phi_3} &= \alpha_2 \alpha_3 e^{i\theta}, \\
   \braket{\phi_3}{\phi_1} &= \alpha_3 \alpha_1 e^{i\theta}, 
\end{align}
where real positive numbers $\alpha_i$ are now defined in terms of the modulus of 
the inner products,  
\begin{align}
  \alpha_1 &\equiv \sqrt{\frac{|\braket{\phi_1}{\phi_2}||\braket{\phi_3}{\phi_1}|}
                        {|\braket{\phi_2}{\phi_3}|} }, \\
  \alpha_2 &\equiv \sqrt{\frac{|\braket{\phi_1}{\phi_2}||\braket{\phi_2}{\phi_3}|}
                        {|\braket{\phi_3}{\phi_1}|} }, \\
  \alpha_3 &\equiv \sqrt{\frac{|\braket{\phi_2}{\phi_3}||\braket{\phi_3}{\phi_1}|}
                        {|\braket{\phi_1}{\phi_2}|} }.  
\end{align} 
Writing $Y_{ij} = \braket{y_i}{y_j}$ as before, we can 
express $d (= \tr YN) $ as follows:  
\begin{align}
  d =1+ \doublebra{A} D \doubleket{A}, \label{eq:dADA}
\end{align}
where $\doubleket{A}$ is the three-component vector defined as 
\begin{align}
 \doubleket{A} = 
  \left(
    \begin{array}{c}
      \alpha_{1}\ket{y_{1}}\\
      \alpha_{2}\ket{y_{2}}\\
      \alpha_{3}\ket{y_{3}}\\
    \end{array}
  \right), 
\end{align}        
and $D$ is the $3 \times 3$ matrix given by  
\begin{align}
D=\left(
  \begin{array}{ccc}
    0 & e^{-i\theta} & e^{i\theta} \\
    e^{i\theta} & 0 & e^{-i\theta} \\
    e^{-i\theta} & e^{i\theta} & 0 \\
  \end{array}
\right). 
\end{align} 
The Hermitian matrix $D$ is readily diagonalized as 
\begin{align}
  U^\dagger D U = {\rm diag} (\lambda_1,\lambda_2,\lambda_3), 
\end{align}
by the unitary matrix given by
\begin{align}
U=\frac{1}{\sqrt{3}}\left(
   \begin{array}{ccc}
     1 & 1 & 1\\
     1 & \omega & \omega^2 \\
     1 & \omega^2 & \omega \\
   \end{array}
\right), 
\end{align} 
where $\omega \equiv e^{i\frac{2}{3}\pi}$ is the cubic root of unity.  
The eigenvalues of $D$ are found to be  
\begin{align} 
 &\lambda_1 \equiv 2\cos\theta,\ \  
  \lambda_2 \equiv 2\cos\left(\theta - 2\pi/3 \right), \nonumber \\   
 &\lambda_3 \equiv 2\cos\left(\theta + 2\pi/3 \right). 
\end{align} 
We can now express $d$ of Eq.~(\ref{eq:dADA}) in the following form: 
\begin{align}
  d = 1 + \frac{1}{3} \Big(  
             \lambda_1 \big|\ket{a_1}\big|^2  
           + \lambda_2 \big|\ket{a_2}\big|^2  
           + \lambda_3 \big|\ket{a_3}\big|^2            
          \Big), \label{eq:d3sum}
\end{align} 
where the three vectors $\ket{a_i}$ are defined by 
\begin{align}
 &\left( \begin{array}{c}
             \ket{a_1} \\
             \ket{a_2} \\
             \ket{a_3} \\
          \end{array} 
  \right)
  = \sqrt{3}\,U^\dagger \doubleket{A} \nonumber \\
 &= 
  \left( \begin{array}{c}
           \alpha_{1}\ket{y_{1}}+         \alpha_{2}\ket{y_{2}}
                                +         \alpha_{3}\ket{y_{3}} \\
           \alpha_{1}\ket{y_{1}}+\omega^2 \alpha_{2}\ket{y_{2}} 
                                +\omega   \alpha_{3}\ket{y_{3}} \\
           \alpha_{1}\ket{y_{1}}+\omega   \alpha_{2}\ket{y_{2}}
                                +\omega^2 \alpha_{3}\ket{y_{3}} \\
         \end{array}
  \right).  
\end{align}

Equation (\ref{eq:d3sum}) shows that $d$ is a linear combination of squared 
norms of the three vectors $\ket{a_i}$. 
Remember that $\ket{y_i}$ is arbitrary with only one constraint on its norm, 
$\big|\ket{y_i}\big|=\sqrt{\eta_i}$. The maximum and minimum of the norm 
of each $\ket{a_i}$ can easily be obtained: The minimum of $\big|\ket{a_i}\big|$ is 0 
if the three lengths $\alpha_i\sqrt{\eta_i}$ satisfy the triangle condition 
(\ref{eq:triangle_relation}). 
Otherwise, the minimum is given by $\alpha_1\sqrt{\eta_1}-
\alpha_2\sqrt{\eta_2}-\alpha_3\sqrt{\eta_3}$, where we assume $\alpha_1\sqrt{\eta_1}$ 
is the greatest of all $\alpha_i\sqrt{\eta_i}$.  The maximum of $\big|\ket{a_i}\big|$ 
is evidently given by $\alpha_1\sqrt{\eta_1}+\alpha_2\sqrt{\eta_2}+\alpha_3\sqrt{\eta_3}$. 
We also notice that the sum of three squared norms of $\ket{a_i}$ is constant.                           
\begin{align} 
  \sum_{i=1}^3 \big|\ket{a_i}\big|^2 = 3\,\doublebraket{A}{A}=
   3 \sum_{i=1}^3 \alpha_i^2 \eta_i.  \label{eq:norm-identity}   
\end{align}
Using this identity, we can rewrite $d$ in terms of any pair of squared norms of 
$\ket{a_i}$, which is still difficult to analytically minimize. However, if 
two of the eigenvalues $\lambda_i$ are equal, $d$ can be expressed by the  
squared norm of a single vector $\ket{a_i}$, which can be minimized very easily. 
We find that this occurs if and only if $\theta=0$ or $\theta=\pi/3$. 
Clearly, these two cases correspond to the case of real $\Gamma$, which was solved in the 
preceding section. It can be verified that we obtain the same results 
as before in these cases. 

There is one instance for which we can easily obtain an analytical result 
even if $\Gamma$ is complex. This is the case in which three $\alpha_i\sqrt{\eta_i}$ 
are equal to each other. 
Eliminating $\big|\ket{a_3}\big|$ by using  
Eq.~(\ref{eq:norm-identity}), we have 
\begin{align}
  d = 1 &+ \lambda_3 \sum_{i=1}^3 \alpha_i^2\eta_i \nonumber \\
        &+ \frac{1}{3} \left( (\lambda_1-\lambda_3) \big|\ket{a_1}\big|^2 
                             +(\lambda_2-\lambda_3) \big|\ket{a_2}\big|^2 \right). 
\end{align}
We find that the two norms $\big|\ket{a_1}\big|$ and $\big|\ket{a_2}\big|$ 
can simultaneously be 0 if $\alpha_i\sqrt{\eta_i}$ is constant. 
This is possible by choosing $\ket{y_i}$, for example, 
to be single-component complex vectors so that three $\alpha_i\ket{y_i}$ make an 
equilateral triangle in the complex plane. 
\begin{align}
  \ket{y_1}=\sqrt{\eta_1},\ \ket{y_2}=\sqrt{\eta_2}\omega^2,\ 
  \ket{y_3}=\sqrt{\eta_3}\omega.
\end{align} 
Note that $\lambda_1 \ge \lambda_3$ and $\lambda_2 \ge \lambda_3$. 
Thus, the minimum of $d$ is given by 
\begin{align}
   d_{\min} = 1 + 2\cos(\theta+\frac{2}{3}\pi) \sum_{i=1}^3 \alpha_i^2\eta_i. 
\end{align}
The attainability condition (\ref{eq:attainability}) requires that 
\begin{align}
  x_i^{\rmR} = 1 + 2\alpha_i^2 \cos(\theta+\frac{2}{3}\pi). \label{eq:xsym}
\end{align} 
The positivity of $N-X^{\rmR}$ can be verified in the following way:
\begin{align}
  & N-X^{\rmR} \nonumber \\ 
  &= \text{diag}(\alpha_1,\alpha_2,\alpha_3) 
        \left( D^* - \lambda_3 \mbold{1} \right) 
     \text{diag}(\alpha_1,\alpha_2,\alpha_3) \nonumber \\
  & \ge 0, 
\end{align}
where $D^*$ is the complex conjugate of $D$, and the inequality follows 
because $\lambda_3$ is also the smallest eigenvalue of $D^*$. 
Therefore, the set of $x_i^\rmR$ given by Eq.~(\ref{eq:xsym}) is the solution of 
the relaxed problem. If all $x_i^\rmR$ are non-negative, they are also the 
solution of the nonrelaxed problem. The maximum success probability is then 
given by $d_{\min}$, 
\begin{align}
   p_{\max} = d_{\min} = 1 + 2\cos(\theta+\frac{2}{3}\pi) \sum_{i=1}^3 \alpha_i^2\eta_i. 
\end{align}   
Otherwise, the problem is reduced to a certain two-state problem.           

This result is a generalization of the symmetric states 
considered in Ref.~\cite{Chefles98}, though the number of states is limited to 
three here. For the symmetric states, the absolute values of all mutual inner products are   
the same, implying a constant $\alpha_i$, and the occurrence probabilities $\eta_i$ 
are assumed equal. In this case, $x_i^\rmR$ are shown to be the same and nonnegative, and 
the maximal success probability is given by 
\begin{align}
  p_{\max} = 1+2\gamma\cos(\theta+\frac{2}{3}\pi), 
\end{align}
where 
$\gamma=|\braket{\phi_1}{\phi_2}|=|\braket{\phi_2}{\phi_3}|=|\braket{\phi_3}{\phi_1}|$. 
    
\section{Summary and concluding remarks}
We have established some general properties of unambiguous discrimination of 
$n$ linearly independent pure states. By formulating the problem as semidefinite 
programming, we have shown that the solution of the problem is unique. 
In unambiguous discrimination, the optimal strategy may produce
a vanishing probability of identifying the input state with some of the states. 
To deal with this possibility, it has been shown to be very convenient to consider a problem 
in which some constraints on the variables are relaxed. 
Applying our method to the case of three pure states, we have obtained 
complete analytic solutions if their mutual inner products are real. 
Unfortunately, our method is not generally applicable to the case of complex inner 
products. However, we have found a class of ensemble of three pure 
states with complex inner products for which we obtain an analytical solution. 

It will also be of interest in future studies
to extend our method to discrimination of three pure states with error margin 
\cite{Hayashi08,Sugimoto09} and to discrimination of three unitary processes 
\cite{Hashimoto10}.

\appendix*
\section{Uniqueness of solutions in the $n$-state problem}
We will show the solution of the $n$-state problem is uniquely determined.
The following two propositions hold:
\begin{theorem}
If $\mbold{x}$ and $\mbold{x}'$ are both solutions of the $n$-state problem, 
then $x_{i_0} = x_{i_0}'$ for at least one $i_0$. 
\end{theorem}

\begin{theorem}
Consider the $n$-state problem with an additional condition: One of the 
components $x_{i_0}$ is fixed to be some value. We assume a solution exists. Then, 
the remaining variables are either 0 or uniquely determined by a certain $r$-state 
problem, where $1 \le r \le n-1$.  
\end{theorem}

With these two propositions, the uniqueness of the solution of the $n$-state 
problem follows by induction on the number $n$ of the states. 
It is evident when $n=1$. We assume that 
the solution is unique in the $r$-state problem for $r \le n-1$.  
Let $\mbold{x}$ and $\mbold{x}'$ be two solutions of the $n$-state problem. 
By proposition (I), we have $x_{i_0}=x_{i_0}'$ for a certain $i_0$. Proposition (II) 
states that the remaining variables are determined by a certain $r$-state problem, 
where $1 \le r \le n-1$. By the assumption, we can conclude $\mbold{x}=\mbold{x}'$.  

Now we present the proof of the two propositions. 
\begin{proof}[Proof of Proposition 1]
If $N-X > 0$, then we can add a sufficiently small positive quantity to any 
component $x_i$ so the condition $N-X \ge 0$ is still satisfied while 
the success probability increases. Therefore, $\det(N-X)=0$ if $\mbold{x}$ 
is a solution. Since a convex linear combination of the two solutions is also a 
solution, we have 
\begin{align}
   \det [ N-\kappa X - (1-\kappa) X' ] = 0\ \ (0 \le \kappa \le 1). 
\end{align}
The coefficient of $\kappa^n$ of this determinant is given by 
\begin{align}
    (x_1'-x_1)(x_2'-x_2)\cdots(x_n'-x_n), 
\end{align}
which should be 0, implying $x_{i_0}=x_{i_0}'$ for a certain $i_0$. 
\end{proof}

\begin{proof}[Proof of Proposition 2]
Without loss of generality, we assume $i_0=n$. We write the condition 
(\ref{eq:e0-positivity}) as  
\begin{align}
    & \Delta - \sum_{i=1}^{n-1} x_i \ket{\tilde \phi_i}\bra{\tilde \phi_i} \ge 0, 
               \label{eq:A-positivity} \\
    & \Delta \equiv \mbold{1}_V - x_n \ket{\tilde \phi_n}\bra{\tilde \phi_n}. 
               \label{eq:A-definition}
\end{align}
The assumption that a solution exists to the problem implies $\Delta \ge 0$.
We first consider the case $\Delta > 0$. The condition (\ref{eq:A-positivity}) 
can be written as 
\begin{align}
   \mbold{1}_V - \sum_{i=1}^{n-1} x_i \Delta^{-1/2}\ket{\tilde \phi_i}
     \bra{\tilde \phi_i} \Delta^{-1/2} \ge 0. 
          \label{eq:A-positivity-2}
\end{align}
The states $\{\Delta^{-1/2}\ket{\tilde \phi_i}\}_{i=1}^{n-1}$ are linearly independent. 
By $V'$, we denote the $(n-1)$-dimensional subspace spanned by them. 
We define the following two sets of $n-1$ states in $V'$:
\begin{align}
   \ket{\phi_i'} &= \frac{Q \ket{\phi_i}}{\sqrt{\bra{\phi_i} Q \ket{\phi_i}}}
                    \ \ (i=1,\ldots ,n-1), \\
   \ket{\tilde \phi_i'} &= \sqrt{\bra{\phi_i} Q \ket{\phi_i}} \Delta^{-1/2} 
                            \ket{\tilde\phi_i}\ \ (i=1,\ldots ,n-1),  
\end{align}
where $Q$ is the projector onto the subspace $V'$. 
It can be readily verified that $\{ \ket{\phi_i'} \}_{i=1}^{n-1}$ and 
$\{ \ket{\tilde\phi_i'} \}_{i=1}^{n-1}$ are reciprocal in the sense that 
$\braket{\tilde\phi_i'}{\phi_j'}=\delta_{ij}$. 
The condition (\ref{eq:A-positivity-2}) takes the form given by 
\begin{align}
   \mbold{1}_{V'} - \sum_{i=1}^{n-1} x_i' \ket{\tilde\phi_i'}\bra{\tilde\phi_i'} \ge 0, 
      \label{eq:1V'_minus}
\end{align}
where $x_i' \equiv x_i / \bra{\phi_i} Q \ket{\phi_i}$. 
The success probability to be maximized can be expressed as 
\begin{align}
    p = \sum_{i=1}^n \eta_i x_i = \eta_n x_n + N \sum_{i=1}^{n-1} \eta_i'x_i', 
     \label{eq:p'}
\end{align} 
where
\begin{align*}
\eta_i' &\equiv \eta_i  \bra{\phi_i} Q \ket{\phi_i} /N, \\
N &\equiv \sum_{i=1}^{n-1} \eta_i  \bra{\phi_i} Q \ket{\phi_i}.
\end{align*}
Thus, maximization of $p$ in Eq.~(\ref{eq:p'}) subject to the conditions 
(\ref{eq:1V'_minus}) and $x_i'>0$ is equivalent to the $(n-1)$-state problem 
for the ensemble $\{\eta_i', \ket{\phi_i'} \}_{i=1}^{n-1}$. 

Next, we consider the case in which $\Delta$ is singular. 
We observe that this happens if and only if $x_n = 1/\braket{\tilde\phi_n}{\tilde\phi_n}$, 
and that $\Delta$ is the projector such that 
\begin{align}                       
      &\Delta \ket{\phi_i} = \ket{\phi_i}\ \ (i=1,\ldots, n-1), \\
      &\Delta \ket{\tilde\phi_n} = 0.                       
\end{align} 
We also find that $x_i$ should be 0 unless $\braket{\tilde\phi_n}{\tilde\phi_i}=0$. 
Without loss of generality, we assume that 
\begin{align}
  \braket{\tilde\phi_n}{\tilde\phi_i} = 
  \begin{cases}
     0 & i=1,\ldots,r \\
     \text{nonzero} & i=r+1,\ldots, n-1, 
  \end{cases}
\end{align}
where $0 \le r \le n-1$. If $r=0$ then all $x_i$ are 0. We assume $r \ge 1$. 
Let $V'_r$ be the subspace spanned by the linearly independent states 
$\{ \ket{\tilde \phi_i'} \}_{i=1}^r$.  Since $V'_r$ is evidently contained in 
the support of $\Delta$, the condition (\ref{eq:A-positivity}) can be written as 
\begin{align}
   \mbold{1}_{V'_r} - \sum_{i=1}^{r} x_i \ket{\tilde\phi_i}\bra{\tilde\phi_i} \ge 0. 
     \label{eq:1V'_minus_2}
\end{align}
The states which are reciprocal to the set $\{ \ket{\tilde \phi_i} \}_{i=1}^r$ 
are not generally normalized. 
However, it is always possible to obtain the set of normalized states 
$\{ \ket{\phi_i'} \}_{i=1}^r$ which is reciprocal to the set 
$\{ c_i \ket{\tilde \phi_i} \}_{i=1}^r$ by choosing the coefficients $c_i$
appropriately.  Note that the resultant states $\ket{\phi_i'}$ are unique 
up to a phase factor. 
Thus, maximizing the success probability subject to the conditions 
(\ref{eq:1V'_minus_2}) and $x_i \ge 0$ is equivalent to the $r$-state problem for 
the states $\{ \ket{\phi_i'} \}_{i=1}^r$ with properly modified  
occurrence probabilities $\eta_i'$. 
\end{proof}


\begin{thebibliography}{99}
\bibitem{Helstrom76}
C.~W.~Helstrom, 
{\it Quantum Detection and Estimation Theory} 
(Academic Press, New York, 1976). 

\bibitem{Holevo82} 
A.~S.~Holevo, 
{\it Probabilistic and Statistical Aspects of Quantum Theory} 
(North-Holland, Amsterdam, 1982).

\bibitem{Chefles00}
A.~Chefles, 
Contemp. Phys. {\bf 41}, 401 (2000). 

\bibitem{Ivanovic87}
I.~D.~Ivanovic,
Phys. Lett. A {\bf 123}, 257 (1987).

\bibitem{Dieks88}
D.~Dieks,
Phys. Lett. A {\bf 126}, 303 (1988).

\bibitem{Peres88} 
A.~Peres,
Phys. Lett. A {\bf 128}, 19 (1988). 

\bibitem{Jaeger95}
G.~Jaeger and A.~Shimony, 
Phys. Lett. A {\bf 197}, 83 (1995). 

\bibitem{Clarke01}
R.~B.~M.~Clarke, A.~Chefles, S.~M.~Barnett, and E.~Riis, 
Phys. Rev. A {\bf 63}, 040305 (2001). 

\bibitem{Hayashi08}
A.~Hayashi, T.~Hashimoto, and M.~Horibe, 
Phys. Rev. A {\bf 78}, 012333 (2008).  

\bibitem{Sugimoto09}
H.~Sugimoto, T.~Hashimoto, M.~Horibe, and A.~Hayashi,  
Phys. Rev. A {\bf 80}, 052322, (2009). 

\bibitem{Chefles98} 
A.~Chefles, 
Phys. Lett. A {\bf 239}, 339 (1998).

\bibitem{Chefles_Barnett98}
A.~Chefles and S.~M.~Barnett, 
Phys. Lett. A {\bf 250}, 223 (1998).

\bibitem{Peres98}
A.~Peres and D.~Terno,
J. Phys. A {\bf 31}, 7105 (1998). 

\bibitem{Sun01}   
Y.~Sun, M.~Hillery, and J.~A.~Bergou, 
Phys. Rev. A {\bf 64}, 022311 (2001).  

\bibitem{Zhang01}
S.~Zhang, Y.~Feng, X.~Sun, and M.~Ying, 
Phys. Rev. A {\bf 64}, 062103 (2001). 

\bibitem{Eldar_IEEE03}
Y.~C.~Eldar,
IEEE Trans. Inf. Theory {\bf 49}, 446 (2003).

\bibitem{Jafarizadeh08}
M.~A.~Jafarizadeh, M.~Rezaei, N.~Karimi, and A.~R.~Amiri, 
Phys. Rev. A {\bf 77}, 042314 (2008). 

\bibitem{Samsonov09}
B.~F.~Samsonov, 
Phys. Rev. A {\bf 79}, 042312 (2009). 

\bibitem{Pang09}
S.~Pang and S.~Wu, 
Phys. Rev. A {\bf 80}, 052320 (2009).  

\bibitem{Rudolph03}
T.~Rudolph, R.~W.~Spekkens, and P.~S.~Turner,
Phys. Rev. A {\bf 68}, 010301 (2003). 

\bibitem{Raynal03}
P.~Raynal, N.~Lutkenhaus, and S.~J.~van~Enk, 
Phys. Rev. A {\bf 68}, 022308 (2003). 

\bibitem{Eldar04}
Y.~C.~Eldar, M.~Stojnic, and B.~Hassibi,
Phys. Rev. A {\bf 69}, 062318 (2004). 

\bibitem{Feng04}
Y.~Feng, R.~Duan, and M.~Ying, 
Phys. Rev. A {\bf 70}, 012308 (2004). 

\bibitem{Herzog04}
U.~Herzog and J.~A.~Bergou, 
Phys. Rev. A {\bf 70}, 022302 (2004). 

\bibitem{Herzog07}
U.~Herzog, 
Phys. Rev. A {\bf 75}, 052309 (2007). 

\bibitem{Zhou07}
X.-F.~Zhou, Y.-S.~Zhang, and G.-C.~Guo, 
Phys. Rev. A {\bf 75}, 052314 (2007). 

\bibitem{Kleinmann10}
M.~Kleinmann, H.~Kampermann, and D.~Bruss, 
Phys. Rev. A {\bf 81}, 020304 (2010).

\bibitem{Vandenberghe96}
L.~Vandenberghe and S.~Boyd, 
SIAM Rev. {\bf 38}, 49 (1996).

\bibitem{Yamashita03}
M.~Yamashita, K.~Fujisawa, and M.~Kojima, 
Optim. Meth. Softw., {\bf 18}, 491 (2003).

\bibitem{Hashimoto10}
T.~Hashimoto, A.~Hayashi, M.~Hayashi, and M.~Horibe, 
Phys. Rev. A {\bf 81}, 062327 (2010).

\end{thebibliography}
\end{document}